\documentclass[12pt]{article}

\usepackage{graphicx}

\usepackage[letterpaper,margin=1in]{geometry}

\renewenvironment{abstract}
	{\quotation}
	{\endquotation}

\date{}

\makeatletter
\renewcommand{\fnum@figure}{\textbf{Fig. \thefigure}}
\renewcommand{\fnum@table}{\textbf{Table \thetable}}
\makeatother

\usepackage{scicite}

\usepackage{url}

\usepackage{amsmath,amssymb,amsfonts}
\usepackage{amsthm}
\usepackage{mathrsfs}
\usepackage{multirow}
\usepackage{algorithm}
\usepackage{algorithmicx}
\usepackage{algpseudocode}
\usepackage{listings}
\usepackage{booktabs}
\usepackage{xcolor}
\usepackage{textcomp}

\newcommand{\abs}[1]{\left| #1 \right|}

\newcommand{\mt}{\widetilde{m}}
\newcommand{\nt}{\widetilde{n}}

\newcommand{\Eq}[1]{Eq.~(\ref{#1})}

\def\scititle{
	Topology-Driven Design of Bianisotropic Metasurfaces Through Knot-Particles
}
\title{\bfseries \boldmath \scititle}

\author{
	Nadav~Goshen$^{1}$,
	Yarden~Mazor$^{1\ast}$\and
	\small$^{1}$School of Electrical and Computer Engineering, Tel-Aviv University, Tel-Aviv 6997801, Israel.\and
	\small$^\ast$Corresponding author. Email: yardenm2@tauex.tau.ac.il
}

\begin{document}

\maketitle

\begin{center}
\textbf{One-Sentence Summary:} Knot-particles enable single-layer bianisotropic metasurfaces allowing complex manipulation of electromagnetic wave through their chiral topology.
\end{center}

\begin{abstract} \bfseries \boldmath
Bianisotropic metasurfaces enable advanced electromagnetic wave manipulation through magnetoelectric coupling. Here, we demonstrate how knot-particles enable single-layer bianisotropic control using their inherent topology. Leveraging their geometric properties, we examine 3D wire configurations characterized by the knot winding numbers (p,q), generating balanced electric and magnetic response. Through multipole analysis we demonstrate efficient polarization rotation with high transmission for different knot-particle topologies. We explore the knot-particle topologies required to achieve perfectly matched polarization rotation and derive simple design rules for the knot parameters without resorting to numerical optimization. The microscopic polarizability tensors and macroscopic susceptibilities reveal the trefoil knot exhibits strong chiral bianisotropic behavior through its magnetoelectric coupling tensor. We implement knot-particle metasurfaces using advanced 3D printing which realizes the full 3D geometry of the wires. We present simplified flat designs suitable for Printed Circuit Board (PCB) fabrication that preserve the essential symmetry that enables the bianisotropic properties.
\end{abstract}

\noindent
Bianisotropic metasurfaces, which exhibit magnetoelectric coupling where electric (magnetic) fields induce magnetic (electric) polarization, have emerged as powerful platforms for controlling wave-matter interactions and spin-orbit phenomena. Among these, chiral bianisotropic metasurfaces, characterized by their broken mirror symmetry, have garnered significant attention across multiple disciplines - from quantum photonics to topological photonics \cite{solntsev2021metasurfaces,lininger2023chirality,deng2024advances}. Their unique ability to induce spin-dependent electromagnetic field distributions \cite{mathur1991thomas,bliokh2015spin,petersen2014chiral,lodahl2017chiral} advances quantum information processing. Beyond the optical regime, bianisotropic metasurfaces have revolutionized electromagnetic wave control at microwave and millimeter-wave frequencies. These surfaces tailor the polarization, phase, and amplitude of electromagnetic waves  \cite{achouri2018design,Grbic_metasurface,zhou2021angular}, enabling asymmetric transmission \cite{TotalAbsorption}, polarization conversion \cite{AchouriPolConv}, and anomalous refraction \cite{li2016simultaneous}. Traditional designs often rely on multi-layer architectures or optimized resonator geometries \cite{pfeiffer2014high,Multilayer_Bian1}, requiring complex optimization procedures to achieve the desired bianisotropic response.
These approaches typically require complex optimization procedures to achieve the desired electromagnetic response.

An alternative approach leveraging topological properties emerges through 
toroidal knot-particles, which naturally possess chiral geometry through their inherent mathematical structure. Their topology is preserved under deformations that do not allow the curve to pass through itself, leading to families of equivalent configurations. An example of such knot-particle is shown in Fig \ref{fig1}A together with the basic torus, and each knot is characterized by two integer winding numbers $p,q$ illustrated in the figure. Crucially, knots can exhibit both geometric and topological chirality - where a knot cannot be superimposed on its mirror image through either spatial transformations or continuous deformations \cite{adams2004knot}. This intrinsic chirality, combined with the knot's three-dimensional structure, naturally enables bianisotropic responses without requiring complex multi-layer designs. In \cite{WernerKnots}, it was shown that a single knot has intriguing radiation characteristics, such as strong directional asymmetry of the radiation pattern, and circular polarization. In \cite{mai2020knotted} it was shown that a single knotted metamolecule has significant optical activity by coupling two waveguides through a $(p,q)=(2,5)$ knot.

Here, we demonstrate how single-layer knot-particle metasurfaces enable electromagnetic control through their unique topological features. These elements exhibit controllable electric, magnetic, and magnetoelectric couplings through their intrinsic topology when arranged in 2D-periodic arrays. Our theoretical framework reveals a fundamental relationship between knot parameters and electromagnetic balance - specifically, configurations satisfying $p=q\pm 1$ achieve optimal electric-magnetic dipole coupling with no need for optimization, using simple, geometry-based, parameter selection rules, and enabling a systematic design of bianisotropic metasurfaces, specifically for perfectly matched polarization rotation. Using multipole analysis validated through full-wave simulations, we demonstrate the physical mechanisms behind this functionality, and show that the trefoil knot ($p=2,q=3$) exhibits strong chiral bianisotropic behavior through dominant diagonal terms in its magnetoelectric coupling tensor. Furthermore, we demonstrate how three-dimensional knotted structures can be systematically translated to two-dimensional implementations while preserving their topological properties essential to their electromagnetic function, providing a pathway for practical fabrication. We demonstrate practical implementation through state-of-the-art 3D printing technology, enabling the realization of complex geometries in their full three-dimensional form, and PCB fabrication of simplified planar configurations. 

\section*{Results}

\subsection*{Formulation}
\begin{figure}[H]
\centering
\includegraphics[width=\textwidth]{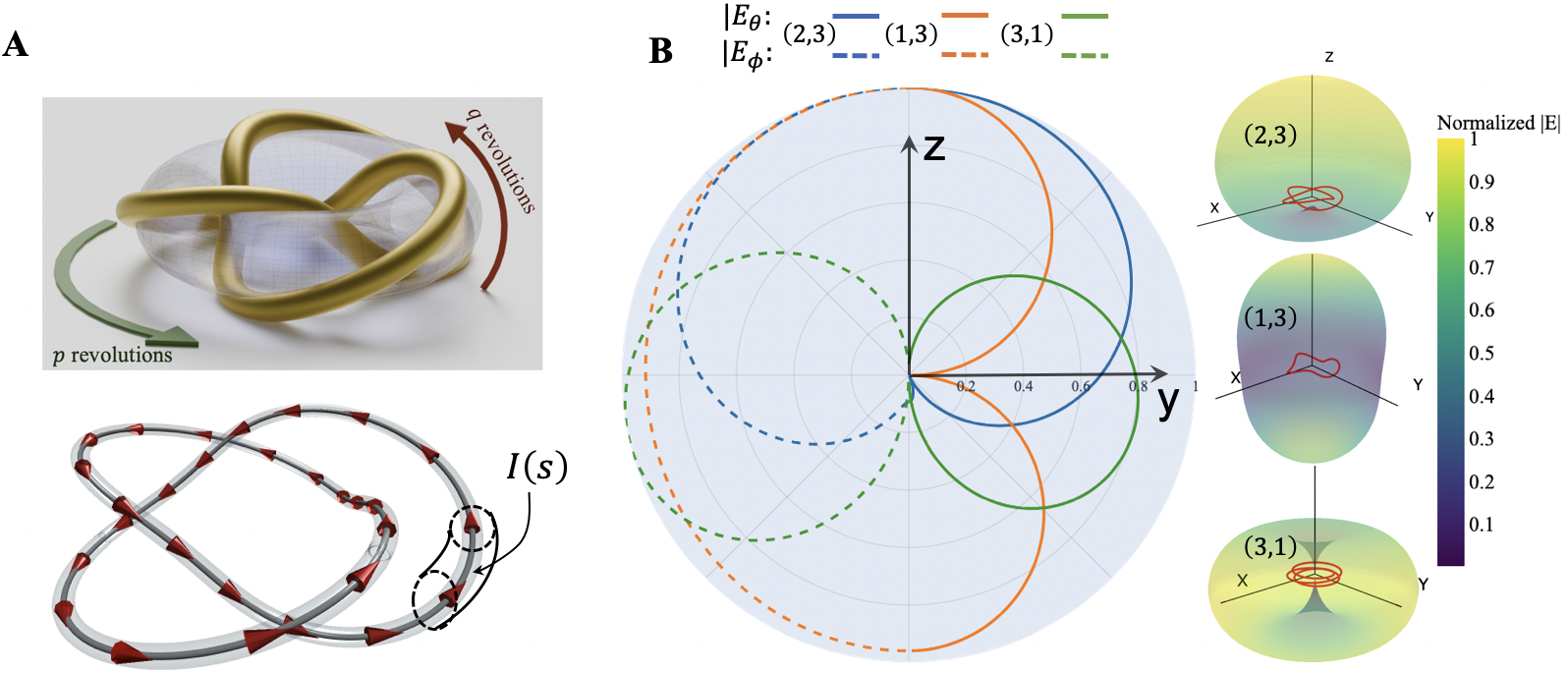}
\caption{\textbf{Toroidal knot-particle geometry and radiation characteristics}. \textbf{A} Top: knot-particle topology on a torus surface, characterized by winding numbers (p,q) representing rotations around the torus and through its core. Bottom: Current distribution $I(s)$ visualization with red arrows indicating current flow along the wire. Knots designed with torus radius $a=2.85$ mm and cross-section radius $b=0.4$ mm for operation at 9 GHz. \textbf{B} Polar plots showing far-field radiation patterns for different knot-particle configurations (p,q): (2,3), (1,3), and (3,1). The left and right sides represent $|E_\theta|$ (solid lines) and $|E_\phi|$ (dashed lines) components, demonstrating distinct radiation characteristics. Right: Corresponding 3D normalized electric field distributions.}
\label{fig1}
\end{figure}

We begin the analysis by taking a look at the basic building block - the toroidal knot. The curve defining the knot is given by \cite{WernerKnots}
\begin{equation}
    \mathbf{r}'(s)=[x'(s),y'(s),z'(s)]=\left[\left(a + b \cos(qs)\right) \cos(ps),\left(a + b \cos(qs)\right) \sin(ps),b \sin(qs)\right]. 
\label{eq:eq1}
\end{equation}
This curve runs on a toroidal surface, with $a,b$ the torus and cross-section radii, respectively. $s\in[0,2\pi]$ is a coordinate parameter. The $p,q$ parameters determine the knot's winding numbers around the cross-section and the origin, effectively controlling the knot's complexity and symmetry properties. A schematic with $p=2,q=3$ is shown in Fig. \ref{fig1}, together with the inscribed torus, and we consider the case where the knot is embedded in a homogeneous material with permittivity $\epsilon_0$ and permeability $\mu_0$. 

The electromagnetic fields radiated by a single toroidal knot can be derived from the vector potential $\mathbf{A}(\mathbf{r})$, by directly integrating over the current distribution along the curve in \Eq{eq:eq1} 

\begin{equation}
    \mathbf{A}(\mathbf{r}) = \mu_0\int_0^{2\pi} I(s) G\left(\mathbf{r},\mathbf{r'(s)}\right) \mathbf{dl},
    \label{eq:Asingle}
\end{equation}
where $\mathbf{dl}=\frac{d\mathbf{r}'}{ds}ds$, is a tangent length element, $G(\mathbf{r},\mathbf{r'})$ is the appropriate Green's function (in free-space $G(\mathbf{r},\mathbf{r'})=\left(4\pi|\mathbf{r}-\mathbf{r}'|\right)^{-1}e^{-jk|\mathbf{r}-\mathbf{r}'|}$), $\mathbf{r}'$ is defined in \Eq{eq:eq1}, and $\mathbf{r}$ is the observation point (throughout this work we consider an $e^{j\omega t}$ time dependence, and suppress it).
From this vector potential, we derive the electric and magnetic fields directly by $\mathbf{E} = -j\omega\mathbf{A} - j\frac{1}{\omega\mu_0\epsilon_0} \nabla(\nabla \cdot \mathbf{A})$ and $\mathbf{H} = \frac{1}{\mu_0} \nabla \times \mathbf{A}$, which exhibit distinct patterns depending on the knot's geometry.

To calculate the radiation characteristics of toroidal knots using Eq. (\ref{eq:Asingle}), we first examine their electric current distribution. As illustrated in Fig.~\ref{fig1}A (bottom), the current $I(s)$ flows along the knot's wire structure, indicated by red arrows along the wire path. Such closed current loops can have many resonance frequencies, which can enable different electromagnetic effects due to different current distributions excited in the knot. In this work, we are interested in the simplest operation regime around the first resonance frequency, where the total wire length is approximately one wavelength \cite{WernerKnots}. The current distribution in this case can be approximated by $I(s) = I_0 e^{js}$, where $I_0$ is the current amplitude (in section "Symmetries and Polarization Rotation Performance of General Knot-Particles" we will provide a deeper intuition for this assumption). This is by no means an exact solution, but it is quite close (as will be shown later), and it provides a lot of intuition into the physical mechanisms and wave phenomena enabled by the knot-particles.

The radiation characteristics of these knots demonstrate strong dependence on their topological parameters $(p,q)$, as shown in Fig.~\ref{fig1}B. The (2,3) configuration, known as the trefoil knot, exhibits pronounced asymmetric radiation along the z-axis (blue line in panel B, top 3D radiation pattern). The (1,3) configuration produces a more symmetric pattern with balanced lobes, while the (3,1) configuration generates a half-wave dipole-like radiation pattern. This diverse behavior arises from the interplay between the knot's topology and its electromagnetic response. The trefoil is a particularly intriguing configuration suited for polarization control applications due to its inherent asymmetric radiation and balanced field components ($|E_\theta|\approx|E_\phi|$).

While the behavior of a single knot-particle is interesting on its own, leveraging the unique scattering properties of knot-particle metasurfaces to control and manipulate the scattered waves requires a planar array. Therefore, we would like extend Eq. (\ref{eq:Asingle}) to account for the field of a 2D array of knot-particles.
Due to the translation invariance of the knot surface, and since we want to consider a plane wave excitation of the surface, the current distribution on each element centered at coordinates $x_{\nt}=\widetilde{n}\Lambda_x, y_{\mt}=\widetilde{m}\Lambda_y$, is taken as
\begin{equation}
 I(s')_{(\widetilde{n},\widetilde{m})}=I(s')_{(0,0)}e^{-jK_x\widetilde{n}\Lambda_x-jK_y\widetilde{m}\Lambda_y},
\end{equation}
 where $K_x$ and $K_y$ are the $x,y$ components of the induced phase gradient of the element excitation (due to an oblique plane-wave excitation, for instance), and $\Lambda_x$ and $\Lambda_y$ denote the periodicities in the $x$ and $y$ directions, respectively. $I(s')_{(0,0)}$ represents the electric current distribution on the element centered at the origin, with the prime notation indicating source coordinates as defined in Eq.~\eqref{eq:eq1}. The radiated fields are derived from the vector potential $\mathbf{A}$ using the 2D-periodic Green's function $G^{P}(\mathbf{r}-\mathbf{r}'(s'))$ (Methods) substituted into Eq. (\ref{eq:Asingle}), which captures the collective response of the periodic array through the sum of all possible spatial harmonics.
The multipole decomposition of a radiating element into an electric and magnetic dipole vectors, $\mathbf{p},\mathbf{m}$ and quadrupole tensors $\mathbf{\mathcal{Q}}^e,\mathbf{\mathcal{Q}}^m$ can provide an "under-the-hood" view of physical mechanisms, and stands as an important link between the discrete element response and representing the surface using various collective and homogenized models. We examine the multipole content of the knot-particles up to quadrupole order, which is necessary due to the complex spatial current distribution in knotted geometries. The relation between the multiples and the current distribution is given in \cite{Multipolar2,Multipolar1}, also summarized in the Methods section.
The calculated multipoles are then utilized to relate the average dipolar and quadrupolar polarization densities and the behavior of the electromagnetic fields on the surface using the multipolar generalized sheet transition condition (GSTC), expressed here for a simplified case of normal incidence \cite{achouri2022multipolar}
\begin{subequations}
\begin{align}
    \hat{z} \times \Delta \mathbf{E} = & - j \omega \mu_0 \mathbf{M}_\parallel 
    + \frac{k_0^2}{2 \epsilon_0} \hat{z} \times \left( \overline{\overline{\mathbf{Q}}}^{\mathrm{e}} \cdot \hat{z} \right) \\
    \hat{z} \times \Delta \mathbf{H} = & j \omega \epsilon_0 \mathbf{P}_\parallel + \frac{k_0^2}{2 \mu_0} \hat{z} \times \left( \overline{\overline{\mathbf{Q}}}^{\mathrm{m}} \cdot \hat{z} \right)
\end{align}
\label{eq:GSTC}
\end{subequations}
where $\Delta \mathbf{E}$ and $\Delta \mathbf{H}$ are the averaged field differences across the metasurface, $\mathbf{M}_\parallel$ and $\mathbf{P}_\parallel$ are the average tangential magnetic and electric polarization densities (per unit area), and $\overline{\overline{\mathbf{Q}}}^{\mathrm{e}}$ and $\overline{\overline{\mathbf{Q}}}^{\mathrm{m}}$ represent the average electric and magnetic quadrupolar surface density tensors, respectively (also per unit area).
The multipole moments of the individual knot-particle derived from Eqs. (\ref{eq:electric_dipole})-(\ref{eq:magnetic_quadrupole}) are related to the surface densities in the GSTC equations (Eq.~\eqref{eq:GSTC}) through division by the unit cell area $A_{\text{unit cell}}$.
The GSTC framework serves a dual purpose in our analysis. On the one hand, it provides a simplified model we can compare against full-wave simulations and assess how accurately this boundary condition captures the complex electromagnetic interactions within the knotted metasurface. On the other hand, these equations enable the extraction of effective polarizabilities ($\overline{\overline{\alpha}}^{ee}$, $\overline{\overline{\alpha}}^{mm}$, and $\overline{\overline{\alpha}}^{em}$) that characterize the electromagnetic response of our structures using the dipole polarizability matrix (and possibly higher order polarizabilities as well). By calculating these polarizability tensors, we can later identify key properties such as intrinsic chirality (indicated by non-zero diagonal terms in $\overline{\overline{\alpha}}^{em}$) and omega-type bianisotropy (revealed by off-diagonal terms).

Using Eq. (\ref{eq:GSTC}), we can relate the scattering parameters to surface currents through generalized sheet transition conditions, which can be expressed using multipole moments \cite{multipole_T_and_R}. This leads to reflection and transmission coefficients for an x-polarized incident field (complete derivation in \cite{supp})

\begin{subequations}\label{eq:FinalRXT}
\begin{align}
R^{xx} &= 
-\tfrac12\Bigl(\tilde{Q}_{xz}^e - \tilde{m}_y + \tilde{p}_x - \tilde{Q}_{yz}^m\Bigr), 
\label{eq:Rxx}\\
R^{yx} &=
-\tfrac12\Bigl(\tilde{p}_y + \tilde{Q}_{xz}^m + \tilde{Q}_{yz}^e + \tilde{m}_x\Bigr),
\label{eq:Ryx}\\T^{xx} &= 
1 \;+\;\tfrac12\Bigl(\tilde{Q}_{xz}^e - \tilde{m}_y - \tilde{p}_x + \tilde{Q}_{yz}^m\Bigr),
\label{eq:Txx}\\
T^{yx} &= 
-\,\tfrac12\Bigl(\tilde{p}_y + \tilde{Q}_{xz}^m - \tilde{Q}_{yz}^e - \tilde{m}_x\Bigr).
\label{eq:Tyx}
\end{align}
\end{subequations}

Here, the superscripts denote co-polarized ($xx$) and cross-polarized ($yx$) components for both reflection ($R$) and transmission ($T$) coefficients. The tilde notation represents normalized multipole moments, with $\tilde{p}=j\omega\eta_0 p/(E_0\Lambda^2_{\text{unitcell}})$, $\tilde{m}=j\omega\mu_0 m/(E_0\Lambda^2_{\text{unitcell}})$, $\tilde{Q}^e={k^2}Q^e/(6\epsilon_0 E_0\Lambda^2_{\text{unitcell}})$, $\tilde{Q}^m={k^2}Q^m\eta_0/(6E_0\Lambda^2_{\text{unitcell}})$, where $\eta_0=\sqrt{\mu_0/\epsilon_0}$ is the free-space impedance, establishing a framework for engineering desired electromagnetic responses through the metasurface's scattering properties.

\subsection*{Trefoil Knot-Particle Metasurface: From Topology to Polarization Control}\label{sec:trefoil}

\begin{figure}[H]
\centering
\includegraphics[width=\textwidth]{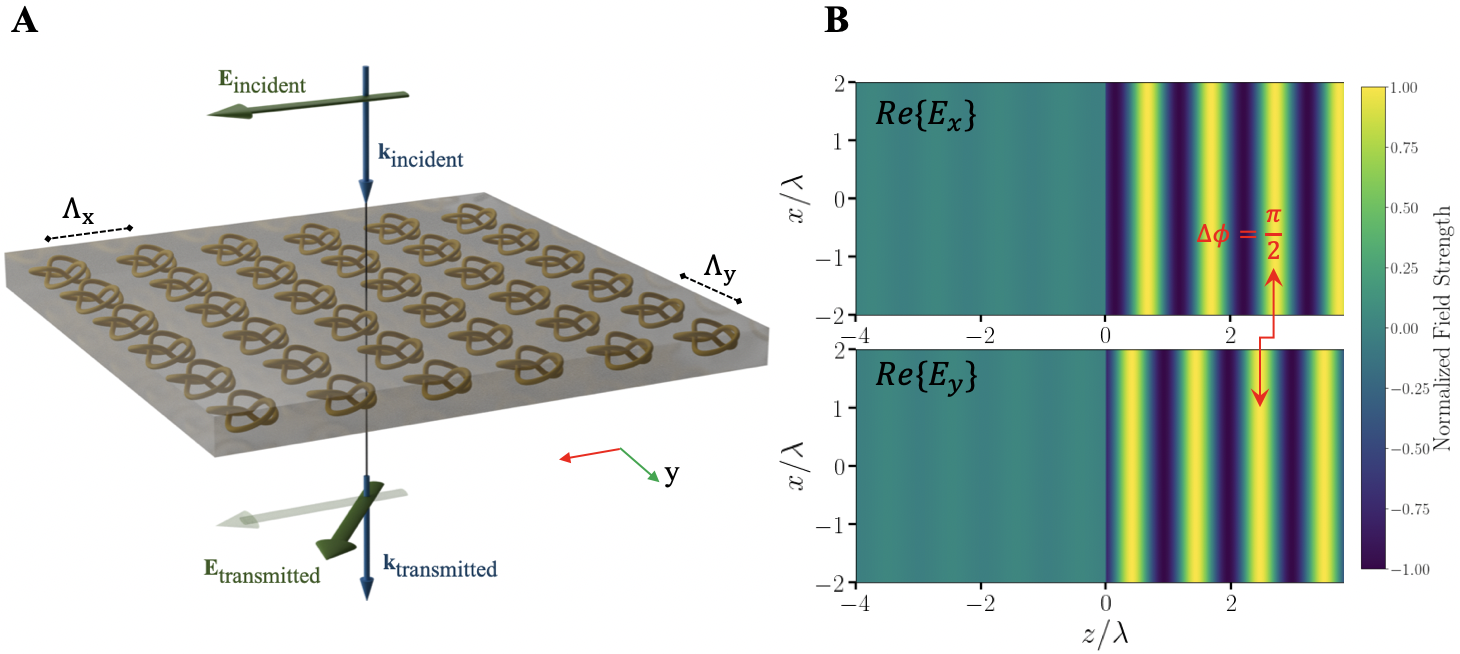}
\caption{\textbf{Trefoil metasurface configuration and radiation characteristics}. \textbf{A} Schematic of the periodic trefoil metasurface with inter-cell dimensions $\Lambda_x = \Lambda_y \leq \lambda_0/2$ under normal incidence excitation, demonstrating the rotation of the transmitted field polarization relative to the incident field. \textbf{B} Real part of x-component (top) and y-component (bottom) of the normalized electric field ($z/\lambda$ vs $x/\lambda$) for infinite array of trefoil-particles (torus radius $a=2.19$ mm, cross-section radius $b=0.8$ mm), showing $\pi/2$ phase difference between components (red arrows) characteristic of circular polarization.}
\label{fig2}
\end{figure}

Among the family of knot-particles metasurfaces, the trefoil configuration ($p=2,q=3$) emerges as a versatile, yet simple design for polarization control, exhibiting remarkable electromagnetic properties that arise from its topology. Starting from the single element radiation characteristics shown in Fig.~\ref{fig1}B (torus radius $a=2.18$ mm, cross-section radius $b=0.4$ mm), we analytically explored parameter variations to enhance the asymmetric radiation pattern, arriving at near-optimized dimensions ($a=2.19$ mm, $b=0.8$ mm). When these elements are periodically arranged ($\Lambda_x = \Lambda_y \leq \lambda_0/2$, Fig.~\ref{fig2}A), this metasurface supports only the fundamental spatial harmonic.
The radiated fields can be calculated using periodic Green's function (Eq.~\eqref{eq:periodicGreen}), where we also incorporate the assumption that the current distribution in each element is $I(s)=I_0e^{js}$ which implicitly assumes operation near the first resonance frequency. 
Fig.~\ref{fig2}B shows the radiated fields, where we see a $\pi/2$ phase difference between the $\hat{\mathbf{x}}$ and $\hat{\mathbf{y}}$ field components, indicating circular polarization. The field distribution also shows remarkable asymmetry along the z-axis, with a 32 dB ratio between forward and backward radiation. Combined with this strong asymmetric radiation (as also demonstrated in Fig.~\ref{fig1}B for a single knot-particle), this enables the structure to function as a perfectly matched polarization rotator with nearly ideal scattering parameters \cite{mai2020knotted} ($R^{xx}=R^{yx}=T^{xx}=0$, $|T^{yx}|=1$).

\begin{figure}[H]
\centering
\includegraphics[width=\textwidth]{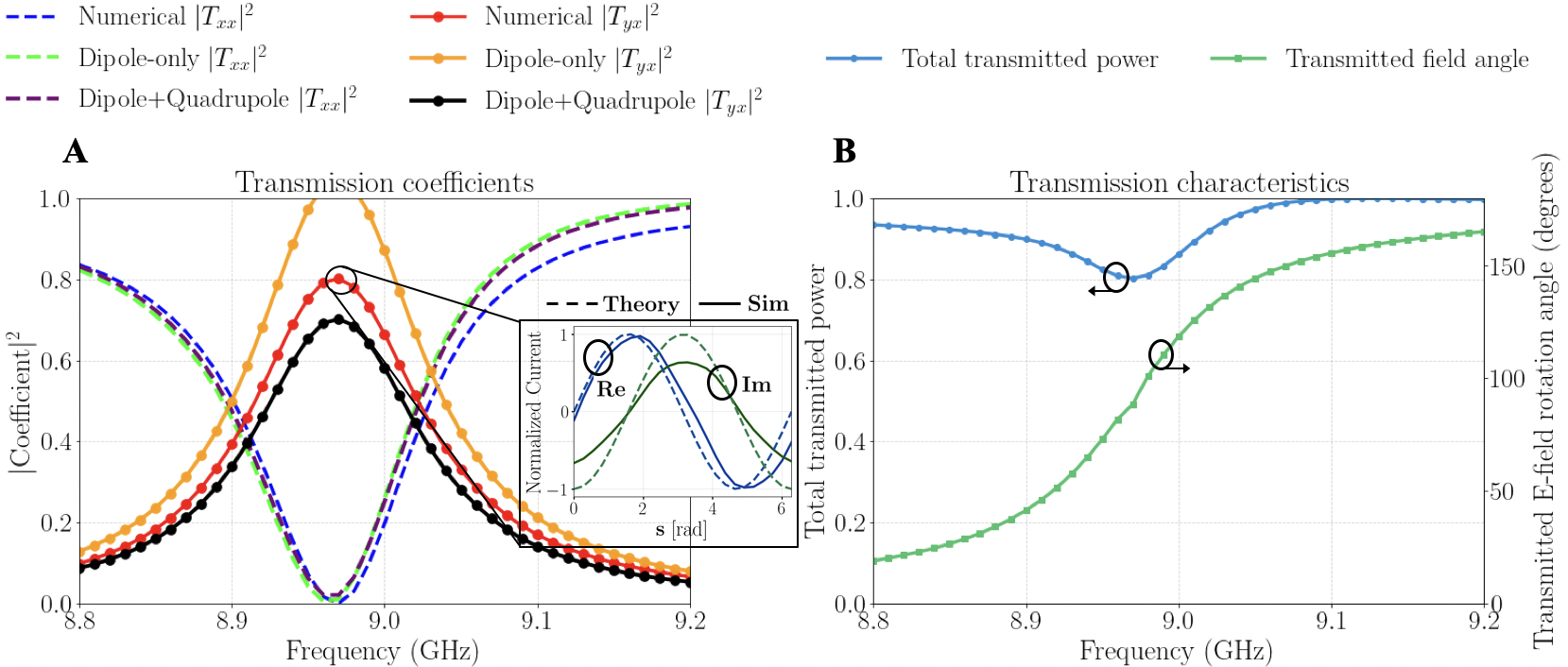}
\caption{\textbf{Transmission analysis of the trefoil knot-particle metasurface}. \textbf{A} Transmission coefficients showing numerical simulation, dipole-only model, and dipole-quadrupole model predictions, with dashed lines for co-polarized ($T_{xx}$) and solid lines with circles for cross-polarized ($T_{yx}$) components. Inset compares normalized current distribution at resonance: simulation (solid lines) versus $I(s)=I_0e^{js}$ approximation (dashed lines), showing real (blue) and imaginary (green) parts. Metasurface parameters: $a=2.19$ mm, $b=0.89$ mm, wire radius $0.1$ mm, $\Lambda_x=\Lambda_y=10$ mm. \textbf{B} Total transmitted power (blue) and field polarization angle (green) as functions of frequency, demonstrating polarization rotation capability.}
\label{fig:3}
\end{figure}

To examine this hypothesis and to obtain a better understanding of the polarization rotation mechanism, numerical simulations using CST Microwave Studio and COMSOL multiphysics were performed. The infinite metasurface, composed of perfect electric conductor (PEC) knot-particles, was modeled using periodic boundary conditions in the xy-plane under normal incidence excitation ($\mathbf{E}^i = E_0\hat{x}$), with the geometric parameters of the knot-particle provided in the caption of Fig. \ref{fig:3}. The current distributions along the knot were extracted and used to calculate the multipole moments through Eqs.~\eqref{eq:electric_dipole}-\eqref{eq:magnetic_quadrupole} numerically, enabling direct comparison between full-wave simulations and the simplified dipole-quadrupole-GSTC model predictions. As shown in the inset of Fig.~\ref{fig:3}A, the normalized current distribution at resonance is close to our analytical approximation. The transmission characteristics shown in Fig.~\ref{fig:3}A demonstrate the metasurface's frequency-dependent polarization control. Numerical simulations \cite{COMSOLref} performed for incident field linearly polarized along $\hat{x}$, show that away from resonance ($f_{res}\approx8.97$ GHz), the co-polarized transmission ($|T^{xx}|$, dashed lines) dominates, preserving the incident polarization. Near $f_{res}$, a transition occurs where the cross-polarized transmission ($|T^{yx}|$, solid lines with circles) becomes dominant, indicating polarization rotation, reaching its maximum value while $|T^{xx}|$ approaches zero, showing efficient polarization rotation at resonance. The analytical predictions through multipole calculations reveal the importance of higher-order contributions. The dipole-only model (orange and green lines) captures the qualitative features well, such as the peak $|T_{xy}|$ frequency, the dip in $|T_{xx}|$, and therefore can be used to study the defining features of knot-particle metasurfaces. However, including the contribution of the quadrupole moments (black line) improves the quantitative predictions,  providing better agreement with numerical simulations (red and blue lines), though some minor deviations remain in the cross-polarized transmission amplitude. This suggests that higher-order multipole contributions play a non-negligible role in the metasurface's response, due to the nontrivial geometry, and a nonquasistatic operation regime. Figure.~\ref{fig:3}B illustrates the practical implications of these transmission characteristics. The transmitted field linear polarization angle (green) exhibits a smooth transition from $20^{\circ}$ to $170^{\circ}$ across the frequency range, demonstrating continuous polarization rotation control. Throughout this operation, the total transmitted power (blue) remains above $80\%$, with maximum efficiency near unity at frequencies away from resonance and a slight dip to approximately $0.8$ at $f_{res}$.

The transmission analysis reveals the importance of multipolar contributions in accurately describing the metasurface response. To gain deeper insight into the underlying electromagnetic operation, we examine the metasurface's response through both microscopic polarizabilities and macroscopic susceptibilities. The microscopic polarizability tensors provide a fundamental understanding of the individual scatterer response, relating the averaged-field differences through~\cite{susceptibility}

\begin{align}
\hat{z} \times \Delta \mathbf{E} &= -j\omega N\overline{\overline{\alpha}}_{\text{mm}} \cdot \mathbf{H}_{\text{t,loc}} -j\omega N\overline{\overline{\alpha}}_{\text{me}} \cdot \mathbf{E}_{\text{t,loc}}, \label{eq:deltaE_local} \\
\hat{z} \times \Delta \mathbf{H} &= j\omega N\overline{\overline{\alpha}}_{\text{ee}} \cdot \mathbf{E}_{\text{t,loc}} +j\omega N\overline{\overline{\alpha}}_{\text{em}} \cdot \mathbf{H}_{\text{t,loc}}, \label{eq:deltaH_local}
\end{align}
where $\overline{\overline{\alpha}}^{\mathrm{ee}}, \overline{\overline{\alpha}}^{\mathrm{mm}}, \overline{\overline{\alpha}}^{\mathrm{em}}$, and $\overline{\overline{\alpha}}^{\mathrm{me}}$ are, respectively, the electric, magnetic, magnetoelectric, and electromagnetic collective polarizability tensors, with the latter two terms representing the collective bianisotropic properties of the metasurface unit cell, and $N$ is the number of scatterers per unit area. The transverse local fields $\mathbf{E}_{\text{t,loc}}=[E_{x,\text{loc}}, E_{y,\text{loc}}, 0]^T$, $\mathbf{H}_{\text{t,loc}}=[H_{x,\text{loc}}, H_{y,\text{loc}}, 0]^T$ must be well defined at each unit cell location. Importantly, the local fields in this description include both the incident field and the interaction fields within the lattice (created by all other knots). To account for the latter, the polarizability matrices can be replaced with \textit{collective} polarizabilities which incorporate the interaction, in which case we exclude the interaction part in the fields on the right side of Eqs. (\ref{eq:deltaE_local})-(\ref{eq:deltaH_local}).

An alternative approach, following~\cite{averagedPolarizability}, models the metasurface through macroscopic surface susceptibilities, enabling a rigorous transition from microscopic to effective medium description

\begin{align}
\hat{z} \times \Delta\mathbf{E} &= -j\omega\mu_0\overline{\overline{\chi}}_{\text{mm}} \cdot \mathbf{H}_{\text{t,av}} -j\omega\sqrt{\epsilon_0\mu_0}\overline{\overline{\chi}}_{\text{me}} \cdot \mathbf{E}_{\text{t,av}}, \label{eq:GSTC_E} \\
\hat{z} \times \Delta\mathbf{H} &= j\omega\epsilon_0\overline{\overline{\chi}}_{\text{ee}} \cdot \mathbf{E}_{\text{t,av}} +j\omega\sqrt{\epsilon_0\mu_0}\overline{\overline{\chi}}_{\text{em}} \cdot \mathbf{H}_{\text{t,av}}, \label{eq:GSTC_H}
\end{align}
where $\mathbf{E}_{\text{t,av}}$ and $\mathbf{H}_{\text{t,av}}$ represent the average transverse fields on both sides of the metasurface. This macroscopic framework proves particularly advantageous for practical implementation in layered dielectric structures, where multiple internal reflections and transmissions occur between interfaces. The susceptibility description characterizes the collective response of the entire structure, avoiding the complexity of tracking individual scatterer interactions in the presence of substrate modes and near-field coupling. For reciprocal metasurfaces, as considered throughout this work, reciprocity constraints apply to both microscopic and macroscopic descriptions, forcing $\overline{\overline{\alpha}}^{\mathrm{ee}}=(\overline{\overline{\alpha}}^{\mathrm{ee}})^{\mathrm{T}}$, $\overline{\overline{\alpha}}^{\mathrm{mm}}=(\overline{\overline{\alpha}}^{\mathrm{mm}})^{\mathrm{T}}$, $\overline{\overline{\alpha}}^{\mathrm{em}}=-(\overline{\overline{\alpha}}^{\mathrm{me}})^{\mathrm{T}}$ and similarly for $\overline{\overline{\chi}}$. These symmetries allow us to fully characterize the electromagnetic response using only the diagonal terms $xx$ (symmetry to $yy$ due to isotropy in the $XY$ plane) and off-diagonal $xy$ (symmetry to $yx$ due to reciprocity) components. Using simulations of bi-directional excitation~\cite{Polarizability}, we extract these microscopic tensors (collective polarizabilities) and map them to surface susceptibilities through generalized sheet transition conditions~\cite{susceptibility,susceptibility2,susceptibility3}, accounting for inter-cell coupling through interaction matrices~\cite{supp}.

\begin{figure}[htb]
\centering
\includegraphics[width=\textwidth]{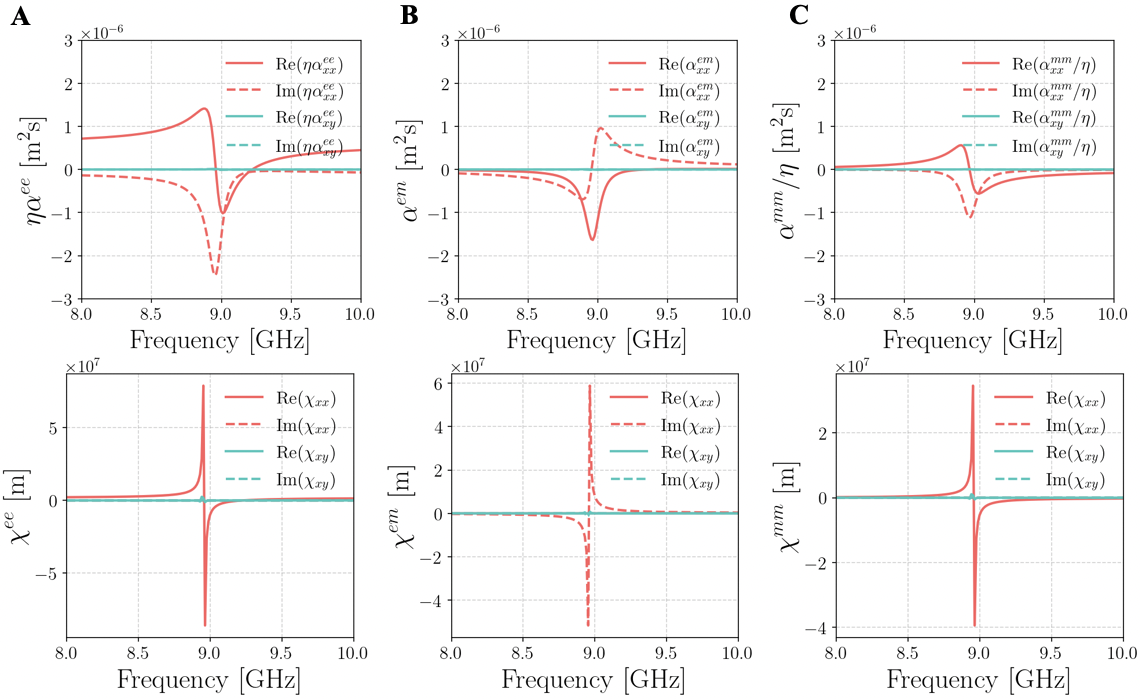}
\caption{\textbf{Electromagnetic response tensors of the trefoil knot-particle metasurface}. Real (solid) and imaginary (dashed) parts of extracted tensors: \textbf{A} scaled electric ($\eta\alpha^{ee}$ top, $\chi^{ee}$ bottom), \textbf{B} magnetoelectric ($\alpha^{em}$ top, $\chi^{em}$ bottom), and \textbf{C} scaled magnetic ($\alpha^{mm}/\eta$ top, $\chi^{mm}$ bottom) responses. The $xx$ and $xy$ components demonstrate strong bianisotropic response near resonance at 9 GHz, with $\chi$ tensors directly measuring the effective surface parameters.}
\label{fig:4}
\end{figure}

Figure.~\ref{fig:4} presents both descriptions: polarizabilities (top) characterizing individual unit cell responses and their corresponding susceptibilities (bottom) describing the effective surface parameters. The polarizabilities demonstrate strong bianisotropic response near resonance, with $\alpha^{ee}$ showing electric dipole excitation, $\alpha^{em}$ revealing significant magnetoelectric coupling through its dominant diagonal components (characteristic of chiral response), and $\alpha^{mm}$ representing the magnetic dipole contribution. Interestingly, the trefoil-particle provides a nearly optimal, single particle chirality around the resonance at $\approx 9$ GHz, with $|\alpha^{em}||\alpha^{me}|\approx|\alpha^{ee}||\alpha^{mm}|$ \cite{supp,PureChiral}. The corresponding susceptibilities reveal similar behavior, with $\chi^{ee}$ exhibiting a sharp resonance ($\sim10^{7}$ m), $\chi^{em}$ maintaining the strong chiral character through its diagonal terms while showing minimal off-diagonal components (indicating negligible omega-type effects), and $\chi^{mm}$ providing the complementary magnetic response ($\sim10^{7}$ m). This balanced coupling between electric and magnetic responses confirms that the key mechanism is the electromagnetic chirality present in our system due to the topological features of the knot-particle, ensuring efficient polarization rotation and impedance matching as demonstrated by the transmission characteristics in Fig.~\ref{fig:3}.

\subsection*{Symmetries and Polarization Rotation Performance of General Knot-Particles}\label{sec:Gen}
The previous sections highlight the performance of a trefoil-particle metasurface as a nearly perfect polarization rotator. One might ask whether or not this performance is specific to the trefoil-particle or if other kinds of knot-particles enable such performance. In Fig. \ref{fig:3}A, we have shown that the current distribution near resonance can be approximated using the simple distribution $I(s)\approx I_0e^{js}$, and Eqs. (\ref{eq:electric_dipole})-(\ref{eq:magnetic_quadrupole}) provide the multipole moments associated with this knot excitation. 

Aside from the obvious simplification achieved by the current distribution assumption, it encapsulates important properties regarding the physics of this problem. In general, the current can be written as $\sum I_ne^{jns}$. The spectral content of each of the current harmonics $I_n$ depends on the excitation, specifically on its expansion on the knot curve (which can be similarly described as $\sum E_ne^{jns}$), and the interaction between the knot parts \cite{WernerKnots}. 
For structures that possess mirror symmetry, and are excited by the normal incident, $\hat{x}$ polarized plane wave, we consider, $I_n=I_{-n}$ is satisfied, rendering a symmetric current distribution. However, in chiral knot-particles (and specifically for the trefoil-particle) an imbalance between $I_n$ and $I_{-n}$ is possible due to the fact that the knot "samples" the exciting field in a chiral manner, which is essential for obtaining the perfect transmission condition (without this imbalance, either the magnetic dipole or electric dipole nullify). When operating in the vicinity of the first resonance (such that the total particle length is $\approx \lambda$) the first current harmonic will tend to be dominant and the assumed current $I(s)\approx I_0e^{js}$ is essentially an idealization of this imbalance, allowing us to examine the optimal performance and focusing on this simplified scenario will allow us to obtain approximate, yet simple design rules for first resonance operation.

\begin{figure}[H]
\centering
\includegraphics[width=\textwidth]{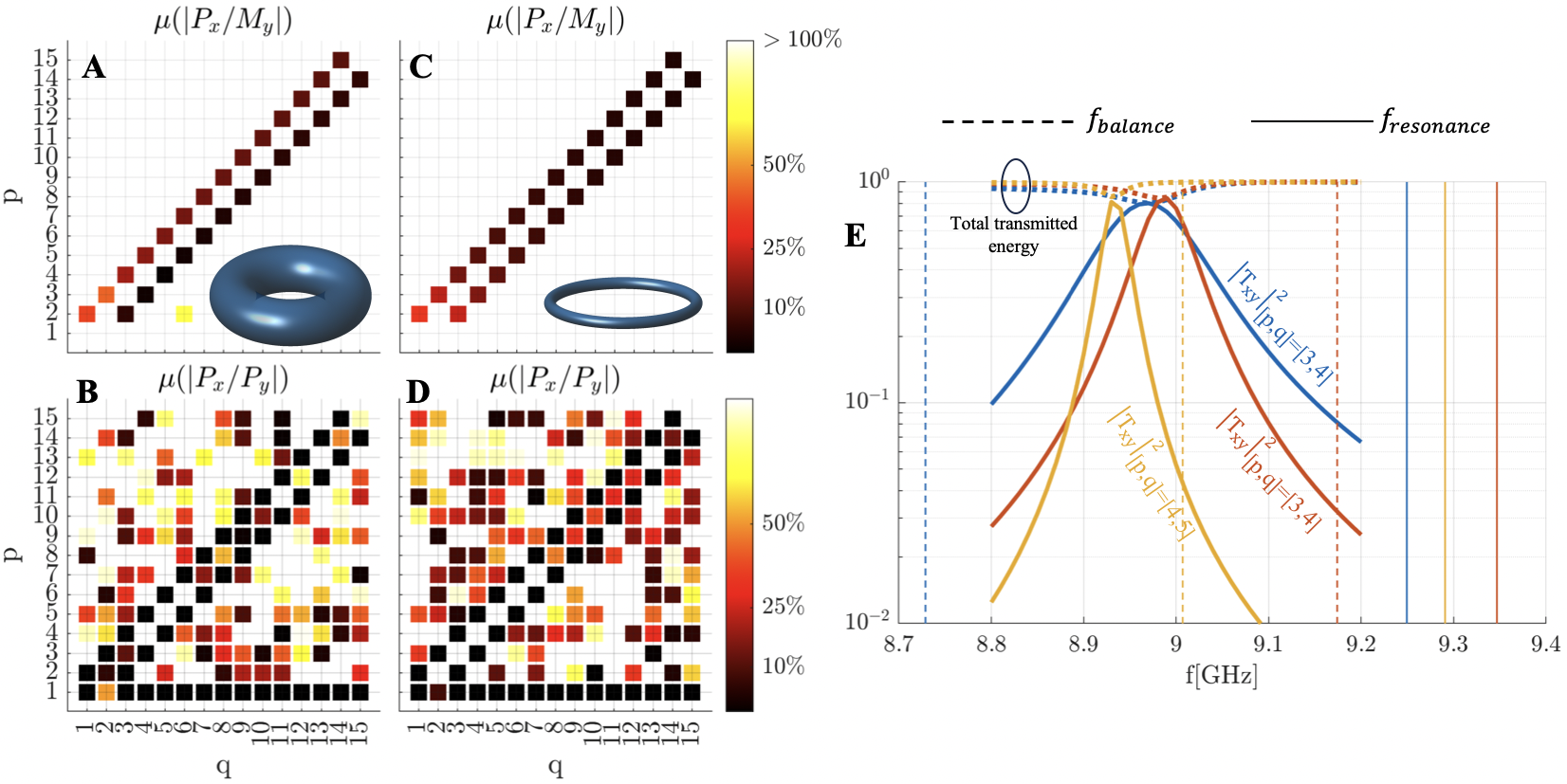}
\caption{\textbf{Electromagnetic balance analysis for different knot-particle configurations.} \textbf{A} Balance score of $r=|p_x/m_y|$ according to $\mu(r)=\abs{(1-r)/(1+r)}$, with the colormap corresponding to a deviation from perfect balance, $r=1$. Deviations larger than $100\%$ are plotted as white squares and therefore are omitted from the plot. \textbf{B} Same as (A), but showcasing the balance of $p_x,p_y$. \textbf{C} and \textbf{D} Same as (A) and (B), but for $a=2.5mm,b=0.2mm$. To calculate (A)-(D), $a,b$ were assumed constant, and for each instance of $(p,q)$, the resonance frequency was calculated through the total length $=\lambda$ requirement. \textbf{E} The transmission curves (obtained using full-wave simulations) for three types of knot-particle metasurfaces, show a higher Q-factor as the knot order becomes larger.}
\label{fig:5_balance}
\end{figure}

If we examine the dipole moments (electric and magnetic), Eq. (\ref{eq:FinalRXT}) tells us that for perfectly matched polarization rotation we must have 
\begin{equation}
    \tilde{p}_{[x,y]}=-\tilde{m}_{[y,x]}, \tilde{p}_x=1, |\tilde{p}_y|=1
    \label{eq:balance_relations}
\end{equation}
While the magnitude of the dipole moments depends on the magnitude of the excited current, the relations between the dipoles provide a fundamental perspective on the suitability of different knot-particle based metasurfaces for perfectly matched polarization rotation. The required dipole moments can be calculated by substituting the approximate current distribution into Eqs. (\ref{eq:electric_dipole})-(\ref{eq:magnetic_quadrupole}). Following that, we can derive an analytical expression for the ratios $\abs{p_x/m_y}$ and $\abs{p_x/p_y}$ to examine whether a certain knot satisfies the required relations. Optimal performance is possible when these relations are as close to $1$ as possible. Figure \ref{fig:5_balance}(A)-(D) shows a visual representation of these relations for various pairs of $p,q$, where the colored squares illustrate the deviation of each ratio from the optimal value (lighter color corresponds to a larger deviation from $ratio=1$, with some reference values given on the color bar. for a certain ratio of $r$ this map is plotted using using the function $\mu(r)=\abs{\frac{1-r}{1+r}}$). Panels (a) and (b) show knots with $a,b$ corresponding to the values used for Fig. \ref{fig:3} (see the inscribed torus in the inset). 
Panels (C) and (D) show the same quantities, this time for $a=2.5mm,b=0.2mm$. 

We can see clearly that while many choices give a roughly balanced $x,y$ dipole response (panels (B,D)), the required magnetoelectric balance is obtained exclusively for $p=q\pm 1$, with $p=q-1$ showing almost perfect balance. Hence, for operation in a simple, first resonance regime, the structure must have a nontrivial chirality ($p=q-1$) to enable the chiral electromagnetic response. Substituting this relation of $p,q$ into Eqs. (\ref{eq:electric_dipole})-(\ref{eq:magnetic_quadrupole}) yields 
\begin{equation}
    \left|\frac{p_x}{c^{-1}m_y}\right|=\frac{2}{\left(2p\pm 1\right)ka}
    \label{eq:balance_px_my},
\end{equation}
which can serve as a guiding design equation to obtain a balanced knot-particle for a certain frequency (balancing $p_y,m_x$ yields a similar relation). Using this equation, the value of $a$ can be extracted (or for a given $a$, the value of the required $k$, and therefore the balance frequency), and then, from the condition that the total knot length equals the wavelength $\lambda$, $b$ is extracted. In addition, we obtain $p_x/p_y=\pm j$, validating the observed $\pi/2$ phase difference between radiated field components seen in Fig. \ref{fig2}, and rendering the source distribution perfectly balanced Huygens sources \cite{shaham2024generalized}.
Qualitatively, both presented cases (Fig. \ref{fig:5_balance}(A,C) and (B,D))   show similar features, which establishes the near-optimal balance as a topological feature. However, we see that quantitatively, the balance performance is worse (the red tint around $[p,q]=[2,3]$ for instance), showing that knots with a significant $z$ dimension have favorable performance.

While many types of knot-particles provide the necessary balance conditions for a near-perfectly matched polarization rotation, they do not have the same performance in terms of other important properties. Figure \ref{fig:5_balance}E shows the transmission of three kinds of knot-particle metasurfaces, $[p,q]=[2,3],[3,4],[4,5]$. All satisfy $p=q-1$ and, therefore, are taken from the optimal diagonal of Fig. \ref{fig:5_balance}(A),(C). We see that as the values of $p,q$ become larger, the resonant peak in transmission becomes narrower, with bandwidths of $1.5\%,0.75\%,0.45\%$, respectively. This can be attributed to the fact that the knot becomes more densely "packed," allowing for larger reactive energy densities to accumulate around the surface, thus increasing the Q-factor. The thin dashed line shows the predicted balance frequency ($f_{balance}$) according to Eq. (\ref{eq:balance_px_my}), and the thin solid lines show the resonance frequency ($f_{resonance}$) according to the total length $=\lambda$ condition. The predictions of these are not the same and present small deviations from the resonance frequency seen in the full-wave simulation results. However, both fall within $<5\%$ of the resonance and can be used as a design guideline to establish most of the required structural parameters.

\subsection*{Realization of Trefoil Knot-Particle Metasurfaces}

\begin{figure}[H]
\centering
\includegraphics[width=\textwidth]{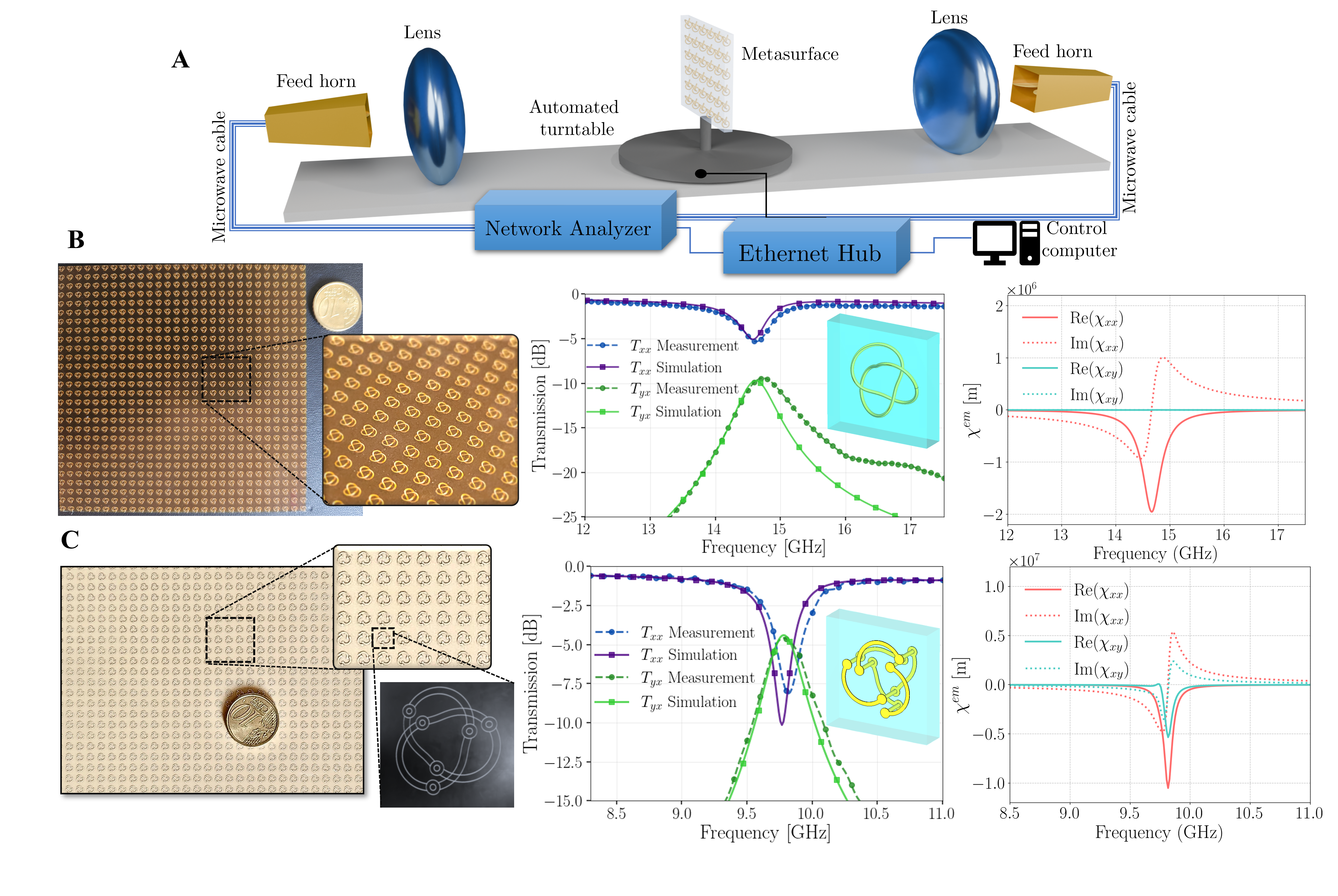}
\caption{\textbf{Experimental validation of knot-particle metasurfaces}. \textbf{A} Free-space measurement setup using the Compass Focused Beam System. \textbf{B} Left: 3D-printed trefoil metasurface using 0.05 mm radius silver ink wire on 0.76 mm thick dielectric substrate ($\Lambda=3.6$ mm, $a=0.85$ mm, $b=0.34$ mm). Middle: Measured and simulated transmission coefficients showing polarization rotation at 14.75 GHz. Right: Retrieved susceptibilities demonstrating chiral bianisotropic response. \textbf{C} Left: PCB implementation with $0.1$mm line width on $D=0.86$mm thick Astra MT77 substrate ($\Lambda=3.9$mm, $a=0.96$mm, $b=0.36$ mm) with inset showing via placement and pad design. Middle: Transmission characteristics with resonance at 9.8 GHz. Right: Retrieved susceptibilities confirming preserved chiral properties. The two implementations are compared to a 10 euro cent coin for scale.}
\label{fig:6}
\end{figure}

To achieve a full 3D geometric realization of the trefoil metasurface design, we fabricated the metasurface using DragonFly IV 3D printer's dual-head system, manufactured by NanoDimension. The numerical validation of the printed structure requires modeling the trefoil element (dimensions detailed in Fig.~\ref{fig:6}B) using the printer material properties: The substrate that hosts the trefoil-particles is a dielectric Ink 1092 substrate ($\epsilon_r = 2.83$, $\tan\delta = 0.021$ at 10~GHz). The conducting knot-particles are composed of AgCite\textsuperscript{\textregistered} conducting nanosilver ink ($\sigma = 2.21 \times 10^7$ S/m) and surface roughness ($R_a<2$ µm). Performing the numerical analysis with these properties reveals preserved polarization rotation functionality, albeit with reduced efficiency due to dielectric losses.

Transmission characteristics were measured using a dual-horn antenna Compass Focused Beam System equipped with lens-focused antennas and automated specimen positioning (Fig.~\ref{fig:6}A, \cite{CompassTech}). Given the fabrication constraint of a maximum 10×10 cm surface size, we designed the metasurface to operate at higher frequencies ($\sim$15 GHz) to achieve an electrical size of $\sim5\lambda_0\times5\lambda_0$ in free space. This increased electrical size ensures sufficient unit cells to represent infinite array behavior while minimizing edge diffraction effects. The fabricated structure is shown in Fig.~\ref{fig:6}B (left). To optimize the fabricated model transmission, while maintaining functionality, we deliberately chose minimal dielectric thickness to reduce material losses in the experimental implementation. Although this design choice impacts the absolute performance (the same design with a lossless dielectric does not provide optimal transmission), numerical simulations of the same system, but with typical low-loss dielectrics ($\tan\delta = 0.002$) demonstrate better transmission performance, albeit still not optimal due to surface roughness inherent to the 3D printing process. (analyzed in detail in Supplementary Materials \cite{supp}). These results confirm that while material losses hinder the performance, maintaining the topological design itself preserves the functionality regardless of the energy absorption. 

The measured transmission characteristics (Fig.~\ref{fig:6}B, middle) show excellent agreement with simulations. The co-polarized transmission ($T_{xx}$) exhibits a resonant dip at 14.5 GHz reaching -5 dB. In comparison, the cross-polarized transmission ($T_{yx}$) shows a peak of -10 dB at $\sim14.6$ GHz, demonstrating the mechanism of polarization rotation predicted by our analysis. The slight frequency shift between the simulation and measurement results ($\sim$ 0.2 GHz) can be attributed to minor fabrication tolerances in the printed structure. Notably, the measured cross-polarized response maintains good agreement with simulations up to 15.5 GHz, with deviations at higher frequencies due to the frequency dispersion of the material parameters. The numerically retrieved electromagnetic susceptibilities (Fig.~\ref{fig:6}B, right) confirm the chiral-bianisotropic nature of the metasurface, with pronounced resonant features in both electric and magnetic components (complete susceptibility analysis available in Supplementary Materials \cite{supp}).

Reducing the losses and achieving better efficiency requires a different fabrication technology, due to the current limitations of 3D printing inks. To demonstrate the potential of low-loss implementation, we realize a semi-planar configuration by essentially guiding the wire over a squared cross-section torus, instead of the round one in Fig. \ref{fig1}A (Supplementary Materials). The projection of the 3D knot onto the xy-plane follows $r'_{2D}(s) = [x'(s), y'(s)] = [(a + b \cos(qs))\cos(ps), (a + b \cos(qs))\sin(ps)]$, and it is printed in parts on the top and bottom sides of the substrate, where the connection is achieved through metalized vias (diameter 0.15 mm) with connection pads (diameter 0.35 mm). We use standard PCB fabrication on Isola Astra MT77 substrate ($\epsilon_r = 3$, $\tan\delta = 0.0017$). Ideally, the projection preserves all symmetries, and the vias are positioned at the points where the original 3D geometry crosses the XY-plane, yielding $z(s)=\frac{D}{2}sgn[\sin(qs)]$, with $D$ being the substrate thickness. In practice, slight offsets are present due to fabrication constraints (see inset in Fig.~\ref{fig:6}C). The surface size is 18$\times$18 cm (corresponding to  $\sim6\lambda$$\times6\lambda$ at the operational frequency). This design enables current flow between layers while maintaining fabrication tolerances, achieving resonance characteristics close to the investigated trefoil topology in the "Trefoil Knot-Particle Metasurface: From Topology to Polarization Control" section.

The fabricated metasurface, shown in Fig.~\ref{fig:6}C (left) with (detailed dimensions in the caption), demonstrates efficient polarization rotation functionality. The transmission measurements (Fig.~\ref{fig:6}C, middle) show excellent agreement with simulations across 8.5-11.5 GHz, with co-polarized transmission ($T_{xx}$) exhibiting a resonant dip at 9.75 GHz reaching -8 dB and cross-polarized transmission ($T_{yx}$) showing a peak of -4.7 dB at 9.7 GHz. The enhanced performance with respect to the 3D printed model stems from the low dielectric losses of the Astra MT77 substrate, which also translates to a narrower resonance peak. Retrieved susceptibilities (Fig.~\ref{fig:6}C, right) confirm that despite geometric simplification, the fundamental chiral-bianisotropic properties are preserved in the planar implementation, as seen in  Fig. \ref{fig:6}C. Unlike the previous case, we also see an additional omega-type response, due to the existence of non-negligible $\chi_{xy}$ values. This is caused by the fact that the offset of the vias from completely symmetric placement makes the "top" and "bottom" parts of the fabricated model a bit asymmetric. However, it is still seen that the chiral behavior is the dominant one, and the polarization rotation functionality is maintained.

\section*{Discussion}

We have demonstrated a novel approach to achieving bianisotropic control through topologically designed knot-particle metasurfaces. Based on the properties of a 2D array of knot-particles, we have shown that periodic structures enhance the single element's asymmetric characteristics through collective response, giving rise to nearly perfect asymmetric fields. Using multipole analysis, we revealed the underlying physical mechanisms driving electromagnetic behavior, showing how resonant knot-particles enable balanced electric and magnetic responses. We have shown that this microscopic picture can be described by both polarizability matrices, and used as a link to a homogenized, macroscopic description susceptibility tensors. Both descriptions point to strong chiral bianisotropic behavior, demonstrated for the trefoil configuration $(p=2,q=3)$. Next, we have shown that all knots from the family $p=q-1$ are naturally suited for the matched polarization rotation functionality by virtue of their topological characteristics, which also provide clear design principles for tailoring knot-particle metasurfaces. We fabricated knot-particle metasurfaces using two techniques - 3D-printed which captures the full 3D geometry of the knot, and planar PCB for a more robust, low-loss implementation. Both designs demonstrated polarization rotation functionality. Despite geometric simplification, the successful translation to standard PCB technology using low-loss dielectrics confirms the practical viability of this approach for single-layer bianisotropic control. It shows that the key properties are the knot-particle chirality and winding numbers $(p,q)$, and the exact geometry can be adapted according to practical considerations. This work establishes knot-particle metasurfaces as a transformative platform for electromagnetic control, enabling vector sensing, polarization manipulation, conformal designs, optical activity, quantum application, and spin-dependent technologies through bianisotropic properties. By unifying topological design with single-layer fabrication, we create a pathway for next-generation electromagnetic devices where geometric principles replace material complexity, fundamentally simplifying the realization of advanced functionalities.

\section*{Materials and Methods}

\subsection*{Radiation from an array}
The 2D periodic Green's function is utilized to calculate the fields radiated by an infinite 2D array of current elements, and is written as a sum of all possible spatial harmonics \cite{munk2005frequency}
\begin{equation}
    \begin{split}
        G^{P}(\mathbf{r}-&\mathbf{r}'(s'))=\frac{\mu_{0}}{2jk_{0}\Lambda_{x}\Lambda_{y}}\sum_{m=-\infty}^{\infty}\sum_{n=-\infty}^{\infty} e^{-jk_{0}(x-x')\left[\frac{K_x}{k_0}+n\left(\frac{\lambda_{0}}{\Lambda_{x}}\right)\right]} \\
        & \cdot e^{-jk_{0}(y-y')\left[\frac{K_y}{k_0}+m\left(\frac{\lambda_{0}}{\Lambda_{y}}\right)\right]}\frac{e^{\pm jk_{z,(n,m)}(z-z')}}{k_{z,(n,m)}/k_{0}},
    \end{split}
    \label{eq:periodicGreen}
\end{equation}
where the longitudinal wavenumber is given by $k_{z,(n,m)} = k_0\sqrt{1-[K_x/k_0+n\lambda_{0}/\Lambda_{x}]^2-[K_y/k_0+m\lambda_{0}/\Lambda_{y}]^2}$, with $\lambda_0$ representing the free-space wavelength, and $(n,m)$ denoting the spatial harmonic orders.

\subsection*{Multipole expansion}
The dipole and quadrupole moments of a general wire current distribution are given by \cite{Multipolar2,Multipolar1}
\begin{subequations}
\begin{align}
p_\alpha &= \frac{1}{j\omega}\left\{\frac{k^2}{2} \int_0^{2\pi}\left[3(\mathbf{r}' \cdot \mathbf{J}_s) r'_\alpha-r'^2 J_{s,\alpha}\right] \frac{j_2(k r')}{(k r')^2} ds + \int_0^{2\pi} J_{s,\alpha} j_0(k r') ds\right\}, \label{eq:electric_dipole} \\
m_\alpha &= \frac{3}{2} \int_0^{2\pi}(\mathbf{r}' \times \mathbf{J}_s)_\alpha \frac{j_1(k r')}{k r'} ds, \label{eq:magnetic_dipole} \\
\mathcal{Q}_{\alpha \beta}^e &= \frac{3}{j\omega} \int_0^{2\pi} \left\{\left[3(r'_\beta J_{s,\alpha} + r'_\alpha J_{s,\beta}) - 2(\mathbf{r}' \cdot \mathbf{J}_s)\delta_{\alpha\beta}\right] \frac{j_1(k r')}{k r'} \right. \nonumber \\
&\quad \left. + \frac{2j_3(k r')}{k r'}\left[\frac{5 r'_\alpha r'_\beta(\mathbf{r}' \cdot \mathbf{J}_s)}{r'^2} - r'_\alpha J_{s,\beta} - r'_\beta J_{s,\alpha} - (\mathbf{r}' \cdot \mathbf{J}_s)\delta_{\alpha\beta}\right]\right\} ds, \label{eq:electric_quadrupole} \\
\mathcal{Q}_{\alpha \beta}^m &= 15 \int_0^{2\pi}\left[r'_\alpha(\mathbf{r}' \times \mathbf{J}_s)_\beta + r'_\beta(\mathbf{r}' \times \mathbf{J}_s)_\alpha\right] \frac{j_2(k r')}{(k r')^2} ds, \label{eq:magnetic_quadrupole}
\end{align}
\end{subequations}
where $\alpha, \beta = \{x,y,z\}$, $j_n(kr')$ is the $n$'th order spherical Bessel function of the first kind, $\mathbf{r}'(s)$ is defined as in Eq.~\eqref{eq:eq1}, and $\mathbf{J}_s = I(s)\hat{\mathbf{t}}(s)$ represents the source current density along the wire with $\hat{\mathbf{t}}(s)$ being the unit tangent vector.

\clearpage
\bibliography{references} 
\bibliographystyle{sciencemag}

\section*{Acknowledgments}
\paragraph*{Funding:}
The authors thank the Israel Science Foundation, Grant 1089/22 for their funding.
\paragraph*{Author contributions:}
N.G. and Y.M. contributed equally to the theoretical development, numerical simulations, and experimental design. Y.M. supervised the project.
\paragraph*{Competing interests:}
The authors declare no competing interests.
\paragraph*{Data and materials availability:}
All data needed to evaluate the conclusions in the paper are present in the paper and/or the Supplementary Materials. Additional data related to this paper may be requested from the authors.


\clearpage

\clearpage

\end{document}